\documentclass[aps,prl,twocolumn,superscriptaddress]{revtex4-2}
\usepackage[caption=false]{subfig}
\usepackage{graphicx,bm}
\usepackage{verbatim,amsmath}
\usepackage{linop}
\usepackage{color,soul}
\usepackage{amsfonts}
\usepackage{ulem}

\newcommand{\ud}{\mathrm{d}}
\begin{document}

\title{Long-Range Order and Quantum Criticality in a Dissipative Spin Chain}

\author{Matthew W. Butcher}
\affiliation{Department of Physics \& Astronomy, Rice University, Houston TX 77005, USA}

\author{J. H. Pixley}
\affiliation{Department of Physics and Astronomy, Center for Materials Theory, Rutgers University, Piscataway, NJ 08854 USA}
\affiliation{Center for Computational Quantum Physics, Flatiron Institute, 162 5th Avenue, New York, NY 10010} 
 \affiliation{Physics Department, Princeton University, Princeton, New Jersey 08544, USA}

\author{Andriy H. Nevidomskyy}
\affiliation{Department of Physics \& Astronomy, Rice University, Houston TX 77005, USA}

\date{\today}

\begin{abstract}
Environmental interaction is a fundamental consideration in any controlled quantum system.  While interaction with a dissipative bath can lead to decoherence, it can also provide desirable emergent effects including induced spin-spin correlations.  In this paper we show that under quite general conditions, a dissipative bosonic bath can induce a long-range ordered phase, without the inclusion of any additional direct spin-spin couplings.  Through a quantum-to-classical mapping and classical Monte Carlo simulation, we investigate the $T=0$ quantum phase transition of an Ising chain embedded in a bosonic bath with Ohmic dissipation.  We show that the quantum critical point is continuous, Lorentz invariant with a dynamical critical exponent $z=1.07(9)$, has correlation length exponent $\nu=0.80(5)$, and anomalous exponent $\eta=1.02(6)$, thus the universality class distinct from the previously studied limiting cases. The implications of our results on experiments in ultracold atomic mixtures and  qubit chains in dissipative environements are discussed.
\end{abstract}

\maketitle

\twocolumngrid

\noindent
\textit{Introduction} -- Decoherence of a quantum two-level system, due to its coupling to the environment, is a key issue in the experimental attempts to improve the stability of a qubit and thus render it more suited to quantum computation~\cite{chuang_quantum_1995, reina_decoherence_2002, shor_scheme_1995, steane_quantum_1998, unruh_maintaining_1995, zurek_decoherence_2008}. Investigating the decoherence of a qubit is thus of obvious import, and its roots in theoretical literature can be traced back to studies of a two-level system in a dissipative environment, referred to as the spin boson model, which has been extensively studied and is a particular limiting case of the Caldeira-Leggett model~\cite{bray_influence_1982, chakravarty_quantum_1982, caldeira_quantum_1983, leggett_dynamics_1987, zurek_pointer_1981, zurek_environment-induced_1982}. If the spin-bath coupling is sufficiently strong, the spin loses its ability to maintain a coherent superposition of ``up" and ``down" states and instead locks into a semiclassical ``localized'' state -- an effect clearly not desirable from the quantum computing perspective. Open quantum systems with bosonic dissipation are not limited in their application to quantum computers, and in fact encompass many experimental endeavors including ultracold atomic gases and ions \cite{bloch_many-body_2008, rubio-abadal_many-body_2019, sabin_impurities_2014, schafer_experimental_2018}.  However, fundamental questions about the nature of decoherence in these complex systems remain unanswered.

When multiple qubits are coupled to the same bath, the dissipation can induce interactions  between distant qubits. This effect is reminiscent of the Ruderman–Kittel–Kasuya–Yosida (RKKY) interaction induced by Friedel oscillations in a Fermi gas \cite{Ruderman_Kittel_1954,Kasuya_1956,Yosida_1957}, although the microscopic mechanism in the presence of a bosonic bath is clearly different.
These boson-induced interactions can allow coherent quantum states to form in a variety of different systems as demonstrated in  trapped ions \cite{cirac_quantum_1995}, superconducting qubits in a microwave cavity \cite{mirhosseini_superconducting_2020,bienfait_phonon-mediated_2019}, and ultracold Bose-Fermi mixtures \cite{desalvo2019observation,bouganne_anomalous_2020, kasper_universal_2020}.

\begin{figure}
    \includegraphics[width=\linewidth]{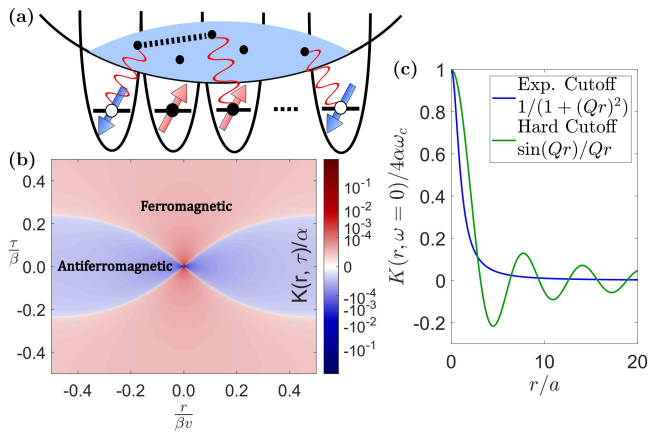}
    \caption{(a)~Diagram of an Ising chain embedded in a BEC bath.  Local confining potentials trap particles in a superposition of two states, which maps to a chain of Ising pseudospins interacting with bath bosons.  (b)~Bath-induced interactions given by Eq.~(\ref{eq:kernel}) in space and imaginary time, in the limit $\omega_c \rightarrow \infty$ ($\lambda \rightarrow 0$), with the color indicating the parity of the interaction.
    (c)~Bosonic RKKY interactions for the cases of exponential and hard cutoff, Eqs.~(\ref{eq:ising_1d_exp}) and (\ref{eq:ising_1d_hard}), obtained by taking the static ($\omega = 0$) limit of the full two-dimensional interactions, here with $Q = 1$.}
    \label{fig:interactions}
\end{figure}

A direct solution of dissipation-induced interactions in qubit arrays that takes into account both their retarded dynamics and long-range nature has remained out of reach.  Instead, theoretical progress has focused on more simplified settings that  either ignore the inter-site dissipation-induced interactions or leave out the dynamical fluctuations of the bosonic medium. For example, arrays of Josephson qubits have been modeled as independent spin-boson systems, leading to locally critical floating phases \cite{tewari_floating_2005, tewari_nature_2006}, which however ignore the bath-induced interactions between qubits.  A problem where these induced interactions 
are expected to play a dominant role, and is central to this work,
is in a 1D spin chain immersed in a bosonic bath. This system, shown in Fig.~\ref{fig:interactions}(a) can be realized using either Bose-Fermi or a Bose-Bose mixture, by placing one atomic species into a deep optical lattice that is embedded in a Bose-Einstein condensate (BEC) \cite{Orth-Lehur2008}.  When the coherence length of the BEC (also known as healing length) is short, this leads to novel universality classes in the presence of short-range inter-qubit interactions \cite{werner_phase_2005, patane_adiabatic_2008}.  In the opposite limit of a very long healing length, one can take the limit of zero lattice spacing, whereby the bath couples to the total value of the spin $S^z=\sum_i s_i^z$, resulting in effectively infinite-range bath-induced interactions and a one-dimensional (1D) Berezinskii-Kosterlitz-Thouless (BKT) transition \cite{winter_quantum_2014}.  In the general case away from these extreme limits, when the full spatial dependence of the bath interactions must be taken into account, only limited progress has been made to date: for instance in the case of only two spins, interesting phenomena such as entanglement \cite{costi_entanglement_2003,McCutcheon2010,Zell2009}, quantum criticality \cite{pixley_pairing_2015,wang_quantum_2021, de_filippis_quantum_2021, zhou_variational_2018}, and coherent dynamics \cite{Orth-Lehur2010, cattaneo_local_2019, karpat_quantum_2020, strathearn_efficient_2018} emerge.  It is clear that the spatial variation of the bath-induced interactions produces nontrivial correlations between coupled spins, however to elucidate their effect on possible ordering and criticality in a spin chain, it is essential to go to the thermodynamic limit, much beyond the two-spin solution, and this has proven to be a very challenging problem.

In this letter, we apply a quantum-to-classical mapping that transforms this one-dimensional quantum problem into a frustrated long-range interacting Ising model in two dimensions, which we simulate using classical Monte Carlo with parallel tempering in coupling constants. Our results demonstrate that a chain of free qubits develops long-range ferromagnetic (FM) order at finite temperature for a sufficiently strong coupling to the bath. We further show that a zero temperature quantum critical point (QCP) separates a quantum paramagnet from the FM phase with unique critical exponents that are distinct from the limit of short-range interactions studied previously~\cite{werner_phase_2005, patane_adiabatic_2008}.  We emphasize that the long-range order arises purely from the spin-spin interactions induced by the dissipative bath, and that the universality class of the QCP is fundamentally changed when accounting for the long-range character of the bath-induced interactions.

\noindent
\textit{Bosonic RKKY Effect} -- The presence of the common bath results in the long-range temporal (retarded) and spatial interaction between the spins, which we refer to as the bosonic RKKY effect by analogy with the RKKY interaction between spins mediated by a Fermi gas.  The model we consider is one of the simplest settings to study the bosonic RKKY effect, which is realized in a chain of spin-1/2 local moments (i.e. qubits) embedded in a common bath of free bosons. The Hamiltonian for this model, which we term the dissipative transverse field Ising model (DTFIM), is
\begin{equation}
\label{eq:hamiltonian}
    H =  - \Delta\sum_{i}\hat{\sigma}^x_i + \sum_{k}\omega_k \hc{b}{k} \op{b}{k} + \sum_{i,k}\hat{\sigma}^z_i e^{ikr_i} g_k\hc{b}{k} + \text{h.c.}
\end{equation}
Here, \(\Delta\) is an applied transverse magnetic field, $\omega_k$ is the dispersion of bath modes, and $g_k$ is the strength of the coupling between local moments and the bath bosons \cite{werner_phase_2005,Orth-Lehur2008}.  The Pauli matrices $\hat{\sigma}_i^x$ and $\hat{\sigma}_i^z$ act on the qubit at position $r_i$, while $\hat{b}_k^\dagger$ and $\hat{b}_k$ create and annihilate, respectively, the bath bosons with momentum $k$.  We stress that in the limit of $g_k=0$ the local moments are completely free and are only coupled to the transverse field, so as to ensure that
any long-range order that may be induced is solely due to the dissipative bosonic bath. The coupling to the bath naturally arises in various settings, perhaps the simplest example being  ultracold Bose--Bose or Bose--Fermi mixtures, where $\omega_k$ denote the low-energy acoustic phonon modes of a bosonic superfluid in which the qubits are immersed (the qubits are represented by the second atomic species that is tightly confined in an optical lattice), see Fig.~\ref{fig:interactions}(a). In this case, the phonon dispersion is linear, \(\omega_k = v|k|\), where $v$ is the sound velocity of the condensate, which we shall assume to be the case for the remainder of this Letter.

By integrating out the bosons and performing a quantum-to-classical mapping \cite{sachdev_quantum_2011}, we arrive at the partition function of a 1+1-dimensional classical Ising model~\cite{Butcher_Pixley_Nevidomskyy_2018}:
\begin{eqnarray}\label{eq:classical_action}
    Z &= &Z_0\sum_{\{s(r,\tau)\}}e^{-S_c}\nonumber\\
    S_c &= -&\Gamma\sum_{i,n}s(i,n)s(i,n+1) \nonumber\\
    &-&\tau_0^2\sum_{i,j}\sum_{m,n}K(r_i-r_j,\tau_m-\tau_n)s(i,m)s(j,n).
\end{eqnarray}
We have introduced the classical Ising variables \(s(i,n)\) which correspond to the eigenvalues of \(\hat{\sigma}_j^z\) evaluated at position \(r_j~\equiv~ja\) along the chain and at imaginary time \(\tau_n~\equiv~n\tau_0\). Here  $Z_0=\prod_k(1-e^{-\beta \omega_k})^{-1}$ is the free boson partition function.  The nearest-neighbor imaginary time interaction \(\Gamma = -\frac{1}{2}\ln(\tanh(\Delta\tau_0))\) that arises from the quantum-to-classical mapping~\cite{sachdev_quantum_2011} does not affect the universality, and we take a constant $\Delta = 1$ without loss of generality. The spatial and imaginary-time dimensions of the system have lengths $L$ and $\beta = N\tau_0$, respectively, with lattice constant $a = 1$.

Similar to the single spin boson model~\cite{caldeira_quantum_1983, leggett_dynamics_1987}, coupling to the bath is captured by a frequency-dependent function, the spectral density $J(\omega) = \pi\sum_k |g_k|^2\delta(\omega - \omega_k)$.  For the case of acoustic phonons in the BEC, the coupling coefficients $g_k$ in Eq.~(\ref{eq:hamiltonian}) scale as $g_k \sim k^{1/2}$~\cite{Orth-Lehur2008}, and the resulting  spectral density is therefore Ohmic, i.e. a  linear function of frequency:
\begin{equation}
    J(\omega) 
    = 2\pi\alpha\omega f(\omega/\omega_c).
\end{equation}
Here \(\alpha\) is a dimensionless parameter characterizing the dissipation strength of the bath.  The cutoff function \(f(\omega/\omega_c)\) depends on the physical setting and must decay to zero as  $\omega$ becomes greater than the bath cutoff frequency $\omega_c$. This cutoff function is often taken to be either smooth or abrupt: 
\begin{eqnarray}
    f(\omega/\omega_c) &=& e^{-\omega/\omega_c} \qquad\qquad \text{``exponential'' cutoff}\nonumber
    \label{eq:e-cutoff}
    \\
    f(\omega/\omega_c) &=& \Theta(1 - \omega/\omega_c) \qquad \text{``hard'' cutoff}
\end{eqnarray}
where \(\Theta\) is the Heaviside step function.

The bath-induced interactions \(K(r,\tau)\) take the form
\begin{equation}\label{eq:kernel}
\!K(r, \tau) = \frac{1}{\pi}\int_0^\infty J(\omega) \cos\Big(\frac{r\omega}{v}\Big)\frac{e^{\omega(\beta - |\tau|)} + e^{\omega|\tau|}}{e^{\beta\omega} - 1}\ud\omega,
\end{equation}
\noindent
whose nontrivial dependence on space and imaginary time is shown in Fig.~\ref{fig:interactions}(b).  Notably, $K(r, \tau)$ can be written in a Lorentz-invariant form by introducing the complex coordinate $z = \tau + \frac{i r}{v}$.  The static limit of the bath interactions $K(r,\omega=0) = \int_{0}^{\beta}K(r,\tau)\ud\tau$
gives a clear definition of the bosonic RKKY effect.  Depending on the choice of the cutoff \(f(\omega/\omega_c)\), the bosonic RKKY interactions are either ferromagnetic
\begin{eqnarray}\label{eq:ising_1d_exp}
K(r,\omega= 0) = \frac{4\alpha\omega_c}{1 + (Q r)^2} \qquad \text{exponential cutoff},
\end{eqnarray}
or oscillating
\begin{eqnarray}\label{eq:ising_1d_hard}
K(r,\omega= 0) = 4\alpha\omega_c\frac{\sin(Q r)}{Q r} \qquad \text{hard cutoff}.
\end{eqnarray}
These two distance dependencies are shown in Fig.~\ref{fig:interactions}(c).
The characteristic momentum \(Q\equiv\frac{\omega_c}{v}\) arises naturally from the introduction of the high-frequency cutoff and is analogous to the Fermi momentum in the fermionic RKKY effect.  In the case of a BEC bath, this momentum can be identified with the inverse of the healing length $Q = \xi_h^{-1}$.  Then, the spatial extent of the bath interactions is fully described by the dimensionless parameter $\lambda = (Qa)^{-1} = \xi_h/a$, where $a$ is the lattice spacing.   In the following we focus on the exponential cutoff in Eq.~\eqref{eq:ising_1d_exp} and leave the hard cutoff, which is harder to converge, to future work \cite{Spin-Boson-II}.

Several limiting cases can be understood from the form of the static interactions in Eq.~\eqref{eq:ising_1d_exp}.  At sufficiently high temperatures such that the transverse field $\Delta \ll k_BT$, the DTFIM maps onto a classical Ising chain with long-range interactions that fall off like $\sim r^{-2}$ at large distances.  This model famously exhibits a finite-temperature BKT phase transition \cite{Thouless_1969} to a long-range ordered ferromagnetic phase as $\alpha$ is increased.  This finite-temperature transition will be explored further in future work \cite{Spin-Boson-II}.  At zero temperature $T = 0$, on the other hand, two limits lend themselves to analytical understanding: (i) the limit $\lambda \rightarrow \infty$ corresponds to the BEC healing length $\xi_h \gg a$ much longer than the lattice spacing, and equivalently $Q\to 0$, where all spins in the chain couple to each other equally strongly and form one large ``superspin" which then behaves like the spin-boson model displaying a  BKT transition~ \cite{winter_quantum_2014}. In the opposite limit (ii) $\lambda\rightarrow 0$, the spins are completely decoupled from one another and the DTFIM maps onto a model where each spin couples to an independent bath. This model has been studied previously, including by the present authors~\cite{werner_phase_2005, Butcher_Pixley_Nevidomskyy_2018}. In this work, we explore the most nontrivial case of finite $\lambda \sim 1$, and show that the resulting QCP has distinct critical exponents from the aforementioned limiting cases.

\begin{figure}
    \includegraphics[width=\columnwidth]{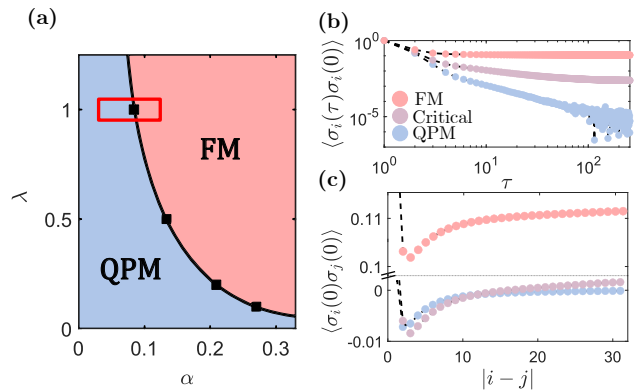}
    \caption{The points shown (a) all belong to the 2nd order universality class described in this text.  Extrapolating off the boundaries to $\lambda \rightarrow \infty$ and $\lambda \rightarrow 0$ results in BKT transitions.  The disconnected correlation functions in the red-boxed region (a) are shown in (b,c), for values of $\alpha = \{0.035, 0.085, 0.09\}$, demonstrating continuously variable power-law decay in the QPM through the critical point, giving way to long-range ferromagnetic order for $\alpha > \alpha_c$.
    }
    \label{fig:phase_diagram}
\end{figure}

\noindent
\textit{Methods} -- We study the DTFIM by performing classical Monte Carlo simulations on the 2D classical Ising model defined by the partition function in Eq.~(\ref{eq:classical_action}).  The long-range interactions include frustration, which causes an exponential slowdown of autocorrelation times in na\"ive Monte Carlo simulations \cite{Troyer_Wiese_2005}.  In order to counteract this effect, the simulations were performed with a combination of Metropolis updates, modified Wolff cluster updates, and parallel tempering updates \cite{Butcher_Pixley_Nevidomskyy_2018}.
The procedure for cluster updates is based on the Luijten-Bl\"ote modified Wolff algorithm for long-range interactions \cite{Luijten_Blote_1995}, however  the presence of mixed-sign interactions (see Fig.~\ref{fig:interactions}b) necessitates a modification where the acceptance probability for adding a given spin to the cluster is calculated from the absolute value of the interaction strength \cite{pleimling_anisotropic_2001, Butcher_Pixley_Nevidomskyy_2018, Supplement}.  In addition to the cluster updates, parallel tempering \cite{earl_parallel_2005} in the dissipation strength $\alpha$ is employed.  Unless otherwise specified, the simulations were performed with 16 replicas at different values of $\alpha$, with all other parameters equal, and the replicas were exchanged on average every 1000 Metropolis steps and 100 Wolff steps, with 10 observable measurements in between each parallel tempering step.  These hyperparameters were found to provide the best autocorrelation times for the systems and observables of interest. 

In the following, we study the total magnetization \mbox{$ m =  [(NL)^{-1}\sum_{i,n}  s(i,n)], $} that we use to compute the Binder cumulant
\begin{equation}
    U_4=1 - \frac{\langle m^4\rangle}{3\langle m^2\rangle^2}
    \label{eq:binder}
\end{equation}
and the disconnected correlation function 
\begin{equation}
    C(|i-j|,\tau) = \langle \sigma_i(\tau) \sigma_j(0)\rangle
\label{eqn:corrfun}
\end{equation}
as probes of the critical properties and relevant phases.
The angle brackets \(\langle\cdot\rangle\) will be used to denote a Monte Carlo average. The imaginary time discretization varies according to $\tau_0 = \omega_c^{-1}$, and should not affect the universality \cite{werner_phase_2005}.

\noindent
\textit{Quantum Critical Point} -- The quantum paramagnet [QPM; the blue region in Fig.~\ref{fig:phase_diagram}(a)] phase occurs for weak dissipation and low temperatures down to \(T = 0\).  As dissipation is increased beyond the critical value $\alpha_c(\lambda)$, the local $\mathbb{Z}_2$ symmetry is broken and the spins order ferromagnetically [FM; the red region in Fig.~\ref{fig:phase_diagram}(a)].  The local self-correlations (Fig.~\ref{fig:phase_diagram}b) and the equal-time spatial correlations (Fig.~\ref{fig:phase_diagram}c) both decay as a power law which varies with $\alpha$ in the QPM phase.  It is interesting that the equal-time correlations are purely antiferromagnetic in the QPM phase, consistent with the form of the interaction shown in Fig.~\ref{fig:interactions}(b), however for $\alpha>\alpha_c$, they approach a positive (ferromagnetic) long-range limit, depicted by red circles in Fig.~\ref{fig:phase_diagram}(c), resulting in the true long-range order with $\langle m\rangle > 0$.

\begin{figure}
    \includegraphics[width=\columnwidth]{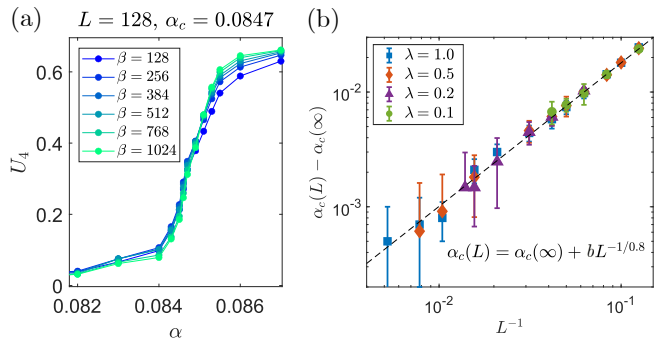}%
    \caption{Calculation of the correlation length exponent $\nu$ for the DTFIM QCP.  $\alpha_c(L)$ is determined by universal Binder cumulant crossings at each fixed value of $L$.  Panel (a) shows $U_4$ for fixed $\lambda = 1.0, L = 128$, with a series of $\beta$, giving $\alpha_c(L = 128) = 0.0847(5)$. For a given $\lambda$, $\nu$ and $\alpha_c(\infty, \lambda)$ are then found by fitting the values of $\alpha_c(L, \lambda)$ to the scaling relation $L^{-1} \sim (\alpha_c(L, \lambda) - \alpha_c(\infty, \lambda))^\nu$, rearranged from Eq.~(\ref{eq:nu_scaling}).  Panel (b) shows the fits for multiple values of $\lambda$ performed simultaneously, giving $\nu = 0.80(5)$.
    }
    \label{fig:nu_fit}
\end{figure}

The long-range interaction makes the finite-size corrections to the critical dissipation $\alpha_c$  significant, so great care is required in extracting the critical exponents
\footnote
{Normally,  the location of the critical point,  as well as the correlation length and dynamical exponent $z$ can be determined by the collapse of the Binder cumulant
$
U_4(\alpha_c,\lambda,\beta,L) = U_4(\alpha_c,\lambda,\beta/L^z)
$
as in Refs.~\cite{werner_phase_2005,sperstad_monte_2010}.  
Unfortunately, 
finite size corrections leads to this procedure failing for this model.
}; see Supplemental Material \cite{Supplement}.
In the following, 
we use the finite size  rounding of the transition  in the limit of zero temperature to extract a strongly $L$-dependent cross-over location $\alpha_c(L,\lambda)$. For clarity of notation the argument $\lambda$ will be suppressed. The finite-size crossovers can be determined by the universal crossings of the Binder cumulant, defined in Eq.~\eqref{eq:binder}, at different system sizes.  For fixed values of $L \in [8, 192]$ at each $\lambda = \omega_c^{-1} \in \{1.0, 0.5, 0.2, 0.1\}$, we determine $\alpha_c(L)$ by extracting the points where $U_4(\alpha, \beta, L)$ lines cross, as shown in Fig.~\ref{fig:nu_fit}a.  The series of $\beta \in \{128, 256, 384, 512, 768, 1024\}$ is fixed, with the number of imaginary time slices $N$ adjusted for different $\tau_0 = \omega_c^{-1}$.  We use the values of $\alpha_c(L)$ from these crossings and later determine the critical exponents directly from the correlation functions.

From a series of $\alpha_c(L)$, the critical dissipation in the thermodynamic limit $\alpha_c$ (specified as $\alpha_c(\infty)$) and the correlation length exponent $\nu$ can be determined by identifying $L$ with the correlation length. The scaling law $\xi \sim (\alpha - \alpha_c)^{-\nu}$ implies the ansatz for the critical coupling
\begin{equation}\label{eq:nu_scaling}
    \alpha_c(\infty) \sim \alpha_c(L) - bL^{-1/\nu}
\end{equation}
for some constant $b$.  Fig.~\ref{fig:nu_fit}b shows a fit to this scaling ansatz for multiple values of $\lambda$, demonstrating a collapse onto a universal scaling law relating $\alpha_c(L)$ and $L$ with $\nu = 0.80(5)$, independent of $\lambda$.
This indicates the entire phase boundary in Fig.~\ref{fig:phase_diagram}a is governed by a common quantum critical universality class with the value of exponent $\nu$ that is distinct from both the case of Josephson junction arrays (limit $\lambda \to 0$) \cite{werner_phase_2005} and from the well-studied case of short-range transverse field Ising model.  These results imply that the long-range dissipative interaction has a profound effect on the universality class of the quantum phase transition.

\begin{figure}
    \includegraphics[width=\columnwidth]{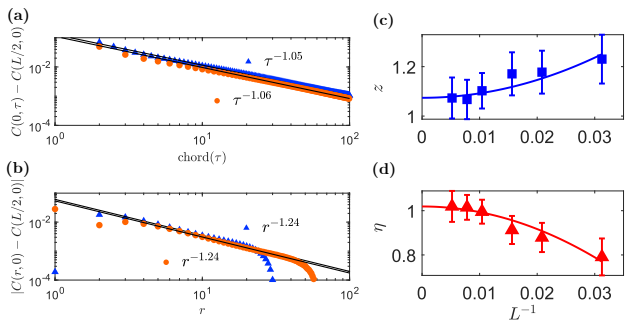}%
    \caption{Connected correlation functions fitted to the scaling form in Eq.~(\ref{eq:correlation_power}) to determine $\eta$ and $z$. The fits show the power-law decay in the (a) same-site and (b) same-time correlation functions for $\lambda=0.5$, $L=64$, $\alpha_c=0.1336$ (blue triangles) and $\lambda = 1.0$, $L = 128$, $\alpha_c = 0.084$ (orange circles), with $\text{chord}(\tau) = \beta/\pi\sin(\pi\tau/\beta)$.  Panels (c) and (d) show $1/L$ extrapolations of $z$ and $\eta$ from the power law fits for $\lambda = 1.0$ at the largest value of $\beta = 2048$.  The results of this extrapolation are $z = 1.07(9)$ and $\eta = 1.02(6)$.}
    \label{fig:z}
\end{figure}

Finally, the correlation functions at the critical point can be used to determine the dynamical exponent $z$, and the anomalous dimension $\eta$.  At the critical point, the connected same-time and same-site correlation functions for $D=1+z$ should follow the universal power law relations \cite{werner_phase_2005}
\begin{eqnarray}\label{eq:correlation_power}
        C(r, \tau=0) - \langle m\rangle^2 &\sim& r^{-(z + \eta - 1)} \nonumber\\
        C(r=0, \tau)  - \langle m\rangle^2 &\sim& \tau^{-(z + \eta - 1)/z},
\end{eqnarray}
where $C(r,\tau)$ is defined in Eq.~\eqref{eqn:corrfun}.
These connected correlation functions are plotted in Fig.~\ref{fig:z}a-b~\footnote{In order to plot the connected correlation function, we subtract the magnetization squared $\langle m\rangle^2$, which we extract from the value of $C(r=L/2, 0)$ on a finite-size-system, relying on the identity $\lim_{r\rightarrow\infty} C(r, 0) = \langle m\rangle^2$.}, and the  finite-size scaling in Fig.~\ref{fig:z}c allows us to extract the critical exponents $\eta = 1.02(6)$ and $z = 1.07(9)$.

\noindent
\textit{Discussion} -- The DTFIM represents a clear-cut example of long-range magnetic order induced purely by environmental bosonic interactions.  By analogy to the fermionic RKKY effect, whose spatial dependence is governed by the UV momentum scale $2k_F$, it is clear that a similar cutoff must appear in the bosonic RKKY. Indeed, the analogous momentum scale is given by the inverse healing length $Q\sim \xi_h^{-1}$ or equivalently the UV momentum cutoff $\omega_c=vQ$, which enters Eqs.~(\ref{eq:ising_1d_exp}) and (\ref{eq:ising_1d_hard}). Despite this UV dependence of the details of the bath-induced interaction, the quantum critical behavior for the exponential cutoff studied in this work remains universal and in particular, does not depend on the finite value of $\lambda \sim Q^{-1}$.

Intriguingly, we find that the quantum critical exponents found here for finite $\lambda$ characterize a novel universality class, fundamentally distinct from the previously studied limits of $\lambda\to \infty$ (i.e. $\omega_c \to 0$)~\cite{winter_quantum_2014} and the limit of $\lambda\rightarrow 0$, which corresponds to each spin in a chain coupled to an independent bath~\cite{ Butcher_Pixley_Nevidomskyy_2018}.  In either of these two asymptotic cases, the quantum critical properties reduce to a BKT transition of the single spin model in a bosonic bath~\cite{bray_influence_1982, chakravarty_quantum_1982,caldeira_quantum_1983}. 

We should add that in the independent bath limit, adding an intrinsic (not bath induced) nearest-neighbor Ising interaction was shown by Werner \textit{et al.} \cite{werner_phase_2005} to lead to a modified dissipative Ising universality class characterized by $z=1.985(15)$, $\nu = 0.638(3)$. By contrast, bond dissipation \cite{sperstad_monte_2010} gives a different set of critical exponents $z = 1.007(15)$, $\nu = 1.005(8)$, the same as the dissipation-free transverse-field Ising model. It is in this context that the present findings are particularly interesting -- we find that with the proper inclusion of long-range bath-induced interactions, the Lorentz-invariance of interaction kernel $K(r,\tau)$ in Eq.~(\ref{eq:kernel}) forces the model to obey conformal invariance with $z = 1$, up to the uncertainty bounds in the present study.  Furthermore, the correlation length exponent takes an anomalous value $\nu = 0.80(5)$.  It is thus clear that the critical properties of dissipative Ising models depend intimately on the details of the bath, and intrinsic length scales therein.

It should be possible to experimentally verify the results of this study, for example, by measuring the susceptibility exponent defined by $\chi \propto |\alpha-\alpha_c|^{-\gamma}$, or the magnetization exponent defined by $\langle \sigma_i^z\rangle \propto |\alpha-\alpha_c|^{\beta}$.  By use of hyperscaling relations, our results predict $\beta \approx 0.4$ and $\gamma \approx 0.8$.  These values are distinct from the values predicted for the non-dissipative 1D transverse-field Ising model ($\beta = 1/8$, $\gamma = 7/4$), as well as those predicted in the examples given above for the limits $\lambda\rightarrow 0$ and $\lambda \rightarrow \infty$.  Since our predicted critical exponents  should apply to the generic case of a finite bath cutoff (a finite BEC healing length), an experimental measurement in, say, ultracold Bose-Fermi mixtures could provide definitive proof of boson-mediated long-range order and the bosonic RKKY effect.

\begin{acknowledgments}
MWB and AHN were supported by the Robert A. Welch Foundation grant no. C-1818. AHN was also supported by the National Science Foundation Division of Materials Research Award DMR-1917511. 
JHP is partially supported by NSF CAREER Grant No. DMR-1941569, and  by the Air Force Office of Scientific Research under Grant No.~FA9550-20-1-0136.
The Flatiron Institute is a division of the Simons Foundation.
All calculations were performed on the Rice University's Center for Research Computing (CRC), supported in part by the Big-Data Private-Cloud Research Cyberinfrastructure MRI-award funded by NSF under grant CNS-1338099.
MWB and AHN acknowledge the hospitality of the Kavli Institute for Theoretical Physics (supported by the NSF Grant No. PHY-1748958), where a portion of this work was performed.
\end{acknowledgments}

\emergencystretch=1em
\bibliography{ms}

\newpage

\begin{widetext}
\section*{Supplementary Material for ``Long-Range Order and Quantum Criticality in a Dissipative Spin Chain''}
\end{widetext}

\section{Cluster updates for mixed-sign interactions}

Due to the mixed ferromagnetic and antiferromagnetic long-range interactions, we employ modified cluster updates originally based on the Wolff algorithm \cite{Wolff_1989}.  The cluster updates can admit long-range interactions \cite{Luijten_Blote_1995} further generalized to mixed-sign interactions \cite{Butcher_Pixley_Nevidomskyy_2018}.  Recently, the authors became aware of a similar method that was also discovered previously \cite{pleimling_anisotropic_2001}, for which this is a generalization.  For a given update step, the cluster begins by selecting a random ``seed'' spin \(s_i\) (here the index \(i\) has absorbed both the spatial index and the imaginary time index for notational simplicity).  The next spin \(s_j\) to be considered for the cluster is selected with probability
\begin{eqnarray}
p_s^i(j) = 1 - e^{-2|\tilde{J}_{ij}|}
\end{eqnarray}
and subsequently added to the cluster if it is favorably aligned with the seed spin \(s_i\), such that
\begin{eqnarray}
p_{add}^{i}(j) = \Theta(\tilde{J}_{ij}s_i s_j).
\end{eqnarray}
In the previous definitions \(\tilde{J}_{ij}\) is the total interaction between spins \(s_i\) and \(s_j\) such that the classical action \(S_c = \frac{1}{2}\sum_{ij}\tilde{J}_{ij}s_i s_j\).  This $S_c$ is the classical action after quantum-to-classical mapping, and in the case of a classical model Hamiltonian $H_\text{cl}$ we can define $S_c \equiv \beta H_\text{cl}$.  We can avoid checking every pair of spins by then calculating the cumulative probability of skipping over a set of \(m-1\) spins and selecting the \(m\)th.  Starting with spin \(s_k\), we skip \(m-1\) spins and select \(s_{k+m}\) with probability
\begin{eqnarray}
C_s^i(k,k+m) = 1 - e^{-2\sum_{j=0}^m |\tilde{J}_{i,k+j}|}.
\end{eqnarray}
A selected spin \(s_m\) is then added to the cluster if it aligned favorably with the seed \(s_i\).  Once all spins have been added or passed over for the seed \(s_i\), another spin in the current cluster is chosen as the seed and the rest of the lattice is queried again by the same procedure.  The cluster building step is only finished once every spin outside the cluster has been rejected by every spin inside the cluster.  Then the spins in the cluster are inverted and the next cluster update begins.

\section{Binder Cumulant Collapse for determination of $\alpha_c$ and $z$}

\begin{figure*}

\includegraphics[width=1.8\columnwidth]{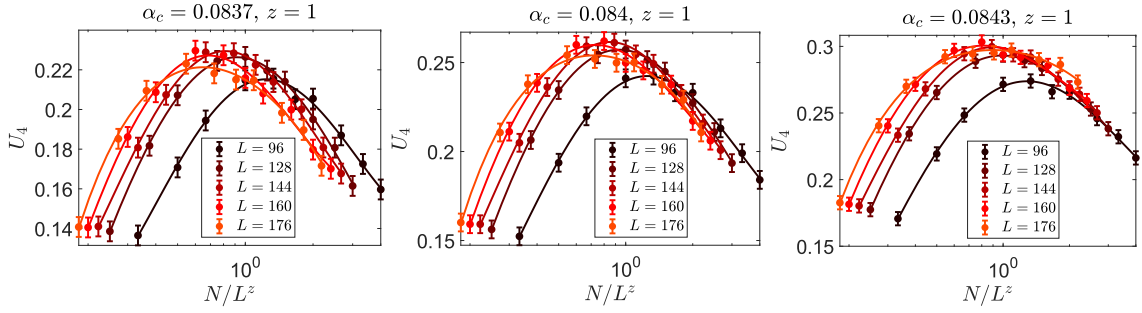}

\caption{Binder cumulant $U_4 = 1 - \frac{\langle m^4\rangle}{3\langle m^2\rangle^2}$ collapse vs. the scaled aspect ratio $N/L^z$ at the error bounds for $\alpha_c = 0.084(3)$ ($\lambda = 1.0$) determined in the main text.  The presumed collapse for $L\rightarrow\infty$ is given with $z = 1$.}

\label{fig:binder_collapse_z}

\end{figure*}

The standard procedure for determining $\alpha_c$ and $z$, as exemplified in Ref.~\cite{werner_phase_2005}, is demonstrated in Fig.~\ref{fig:binder_collapse_z}.  The scaling ansatz for the Binder cumulant $U_4 = 1 - \frac{\langle m^4\rangle}{3\langle m^2\rangle^2} \sim f(\alpha=\alpha_c, \beta/L^z)$ implies that for fixed $\alpha = \alpha_c$, $U_4$ should collapse to a universal function of the aspect ratio $N/L^z$ (where $N = \beta/\tau_0$ according to the quantum-classical mapping).  The strong finite-size effects induced by the long-range interactions in the DTFIM cause a significant drift in the value of the effective critical point $\alpha_c(L)$ for nearly all accessible system sizes.  Therefore, this method is ineffective for determining $\alpha_c$ or $z$ without prior knowledge.

By visual inspection, it is clear there is no collapse whatsoever for values of $\beta \lesssim L$.  This is a result of the shift from 2nd order QCP to 1D BKT at finite temperatures.  On the side of $\beta \gtrsim L$, the collapse is best for small values of $L$ at $\alpha_c = 0.0843$, and improves for the largest values of $L$ at $\alpha_c = 0.0840$, as determined in the main text.  This method is not sufficient for a non-biased determination of $z$ but does support the conclusion $z=1.07(9)$.

\section{Binder Cumulant Crossings for Determination of $\alpha_c(L)$}

\begin{figure}

\includegraphics[width=\columnwidth]{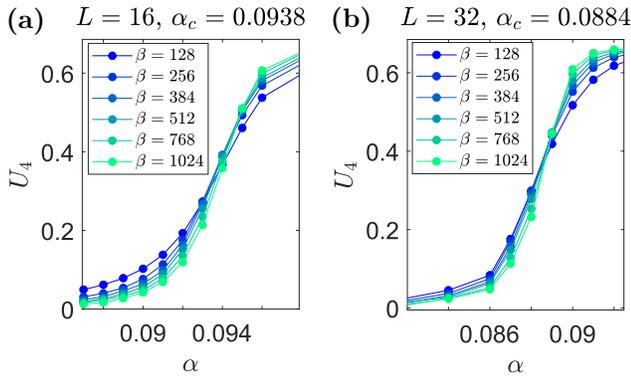}

\caption{Binder cumulant crossings for additional values $L = 16$ (a) and $L = 32$ (b).}

\label{fig:binder_crossing}

\end{figure}

Binder cumulant crossings for some small values of $L$ are displayed in Fig.~\ref{fig:binder_crossing}.  The critical Binder cumulant is near $U_4(\alpha_c) \approx 0.45$ for all $L$ in the limit $\beta\rightarrow \infty$.  This is further evidence that second-order universality is already reached in small system sizes, and the finite-size-dependence of $\alpha_c$ is a result of the power-law scaling of correlation length as described in the main text.

\section{Finite-Size Extrapolation of Correlation Function Fits}

\begin{figure}

\includegraphics[width=\columnwidth]{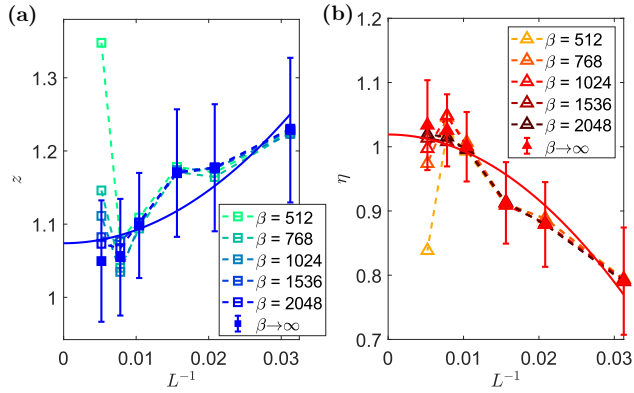}

\caption{Finite-size values for (a) $z$ and (b) $\eta$ are shown overlaid with the $\beta\rightarrow\infty$ extrapolations, which are pictured in the main text.  Error bars are bootstrapped from the fit errors in the correlation function fits, through the $\beta\rightarrow\infty$ and $L\rightarrow\infty$ extrapolations successively.  The finite-size estimates do not depend strongly on $\beta$ except in the case $L=192$ (leftmost points).}

\label{fig:z_eta_extrapolation}

\end{figure}

Full data for correlation function exponent extrapolations is shown in Fig.~\ref{fig:z_eta_extrapolation}.  Correlation functions for all system sizes were fit to the scaling forms shown in the main text to obtain a finite-size estimate for $z(L,\beta)$ and $\eta(L, \beta)$ with $\lambda = 1$.  Then, for each fixed value of $L$, the $\beta\rightarrow\infty$ estimate is obtained by a cubic fit vs. $\beta^{-1}$ according to the ansatz $x(L, \beta) = x(L) + a\beta^{-2} + b\beta^{-3}$, where $x$ is either $z$ or $\eta$.  Finally, the $\beta\rightarrow\infty$ estimates are extrapolated to infinite system size by a fit to the ansatz $x(L) = x + cL^{-2}$.  The cubic term is left off the $L$ extrapolation to prevent overfitting for the small number of sample points.

\end{document}